# Incorporating Graph Attention Mechanism into Geometric Problem Solving Based on Deep Reinforcement Learning

Xiuqin Zhong[a,1], Shengyuan Yan[a,2], Gongqi Lin[a,3], Hongguang Fu[a,*], Liang Xu[a], Siwen Jiang[a], Lei Huang[a], Wei Fang[b,*]

[a]*University of Electronic Science and Technology of China, 611731, Chengdu, China*
[b]*West Virginia Clinical and Translational Science Institute, Morgantown, WV 26506, America*

**Abstract**

In the context of online education, designing an automatic solver for geometric problems has been considered a crucial step towards general math Artificial Intelligence (AI), empowered by natural language understanding and traditional logical inference. In most instances, problems are addressed by adding auxiliary components such as lines or points. However, adding auxiliary components automatically is challenging due to the complexity in selecting suitable auxiliary components especially when pivotal decisions have to be made. The state-of-the-art performance has been achieved by exhausting all possible strategies from the category library to identify the one with the maximum likelihood. However, an extensive strategy search have to be applied to trade accuracy for efficiency. To add auxiliary components automatically and efficiently, we present deep reinforcement learning framework based on the language model, such as BERT. We firstly apply the graph attention mechanism to reduce the strategy-searching space, called AttnStrategy, which only focus on the conclusion-related components. Meanwhile, a novel algorithm, named Automatically Adding Auxiliary Components using Reinforcement Learning framework (A3C-RL), is proposed by forcing an agent to select top strategies, which incorporates the AttnStrategy and BERT as the memory components. Results from extensive experiments show that the proposed A3C-RL algorithm can substantially enhance the average precision by 32.7% compared to the traditional MCTS. In addition, the A3C-RL algorithm outperforms humans on the geometric questions from the annual University Entrance Mathematical Examination of China.

*Keywords:*
Elementary mathematics, Auxiliary components, Attention network, Reinforcement learning, Automated Mathematical Reasoning

## 1. Introduction

Automated mathematical reasoning is a core question of Artificial Intelligence (AI) that dates back to the early days of computer science [1]. Many attempts have been made to design various systems for education. Recently, making the computer pass entrance examinations at different levels of education has surfaced as an important AI challenge [2]. With the rise of online education, automated mathematical reasoning plays an increasingly important role. Automated mathematical reasoning provides interactive and cognitive learning modes for online education. This online interactive learning mode provides students with clear problem-solving solutions and improves their ability to solve problems.

In mathematical reasoning, geometrics by nature is an ideal candidate for pure logical reasoning processed through AI. Actually, several problem-solving systems have been successfully built. For example, Project Aristo [3] showcased a challenging task of improving the performance of AI modeling and reasoning in solving elementary school science and math exam problems. In spite of the progress that has been made, adding geometric auxiliary components such as points or lines, which is common practice in solving geometric problems, is still challenging in AI and automatic reasoning. There are two main challenges for solving geometric problems automatically.

First of all, the full connection (*e.g.*, connection all points in the graph), was applied to search possible solutions, but failed to solve some specific geometric problems. For example, the problem needs to add auxiliary points, and these points do not exist in the original graph. To illustrate this challenge, Figure 1 was created which shows a geometric question (Example 1) in a mathematics test. The geometric question (Example 1) presents as "In Pyramid $P-ABCD$, $PC \perp planeABCD$, $AB \parallel DC$, $DC \perp AC$. Assuming point $E$ is the midpoint of $AB$, is there a point $F$ on edge $PB$ such that $PA \parallel planeCEF$ ?". The proof question in Figure 1 is difficult without adding a new auxiliary mid-point $F$ on the segment $PB$. After setting up mid-point $F$, it needs to connect auxiliary lines $CF$ and $CE$ to obtain the plane $CEF$. Thus, the question cannot be solved without adding the new point $F$. In other cases, some problems need to extend the segment in order to obtain the solution. However, the extension cannot be reachable only using full connection.

Furthermore, a key challenge of adding auxiliary components is strategy selection: selecting relevant strategy that are useful for proving a given conjecture or solving a geometric problem. Solving a geometric problem is essentially a search problem with the goal of finding a sequence of deductions leading from presumed facts to the given conjecture. However, as this traditional way needs to exhaust all possible combinations of points or lines to find the possible solution, this approach is computationally expensive and time-consuming. As a consequence, combination explosion would happen if all connections are simultaneously applied to complex questions. We will illustrate this issue in complex questions in our experiment. The space of this search is combined explosion with today's large mathematical knowledge bases[4, 5], the search can quickly explode beyond the capability of the system, despite the fact that often only a small fraction of facts in the knowledge base are relevant. Strategy selection thus plays a critical role in narrowing down the search space and making it tractable.

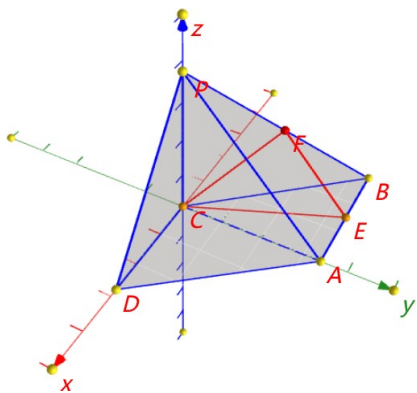

Figure 1: An Example of Geometric Problem with a new plane *CEF* as an auxiliary component

Due to success in computer visions and natural language processing, neural networks have recently provided an efficient way to guide solving mathematical problems automatically, such as theorem proving[6, 7, 8, 9], and demonstrating approximate mathematical reasoning abilities in latent space [10]. In this paper we propose a novel deep reinforcement learning framework with language model, such as BERT [11], which aims at tackling the drawbacks mentioned above.

The key idea of our approach is to represent strategies as a sequence and embed them into vector space. This is different from prior work [12], which proposed a reverse reasoning method with frustration tree to check whether it was possible to generate a proof tree. Our approach is motivated by the observation that whether solving the geometric problems by adding auxiliary components, which can be represented as a sequence that encodes graph information by the strategy change. Then We trained our observation data using BERT to predict the strategy selection. To reduce generating strategies, we applied the graph attention network for strategy selection. Based on the selected strategies, we build a sequence of systems, adding Monte-Carlo tree search [13], and reinforcement learning [14] to solve the geometric problems efficiently and automatically. Our contributions can be summarized as follows:

1) We build strategy networks to reduce the searching space using graph attention mechanism, called AttnStrategy.

2) This research presents a novel algorithm, called Automatically Adding Auxiliary Components using Reinforcement Learning framework (A3C-RL), which incorporates AttnStrategy and deep learning contribution model as the memory components.

3) We use the traditional MCTS mechanism to solve geometric problems by adding auxiliary components, which can be up to 50.8% in accuracy rate. We use it as a baseline for our model.

4) Compared to the traditional MCTS method, the proposed A3C-RL framework improves the accuracy rate of geometric problem-solving by 32.7%, up to 83.5%.

## 2. Related Works

As we have noticed, it is the recent rapid progress in language understanding and generation capabilities [11, 12, 15, 16, 17]. Language modeling using Transformers [15] has been hugely successful for applications like translation and text generation. Improvements made from language modeling have been demonstrated from better pre-training tasks, using various objectives such as auto-regressive generation [12, 16, 17], token masking [11] and sequence masking [18], and algebraic word problems [19, 20]. Recently, Lample and Charton [21] successfully applied Transformers to anti-derivative calculus and solving differential equations, hinting that Transformers were capable of generating the exogenous terms involved in the substitutions required for successful symbolic integration. The Universal Transformer [22], a Transformer with tied weights, was also shown to be successful at more algorithmic tasks. Also, Saxton [23] evaluated the Transformer architecture on a variety of mathematical problems. In our research, instead of training language models on formal mathematics, we trained our model on the sequence of graph features by the strategy-selection task. The strategy-selection task is a specialization of the skip state task that only focuses on the current state.

Before conducting the strategy selection, we need the strategy network to generate all possible strategies based on the given information. Our strategy network includes three main layers according to their functions: the method strategy, the completeness strategy, and the simplification and evaluation strategy. The first layer is the method strategy, which introduces unified methods, equivalent transformation of the conclusion, and auxiliary components. A unified method regards the conclusion as a known condition. For example, to prove: "$a = b$", a unified method uses "$a = b$" as a known condition. The second layer is a completeness strategy, which mainly handles different branching discussions for a problem with the same scenario. Theoretically, all candidate strategies need to be considered and discussed to solve one problem completely. The last layer is the simplification and evaluation branching strategy, which mainly reduces the difficulties of a problem using feature-based skills. This layer transforms the known conditions into equations and solves the selected equations using various combination strategies. In our research, the candidate strategies for the problem are generated by the strategy network with mentioned three layers.

Theoretically, there exists an optimal value function as the strategy network is unconstrained, and we need to search for it. However, implementing strategies for adding auxiliary components in search of an optimal value is a double-edged sword in that while strategies might assist in knowledge base reasoning, inappropriate or excessive use of strategies would overwhelm knowledge base. The AlphaGo team [11] introduced a value network capable of directly estimating the value of state (or a win rate) for any given game position in Go in order to better approximate the value of leaf-node states during the tree search. Hence, to minimize the abuse of strategies and optimize the search, we applied the value network to evaluate the search process with three main steps: state space comparison, conclusion relevance, and evaluation function.

In particular, the state space comparison has a regular experience to add auxiliary components based on computing methods which have the same type of conclusions. The conclusion relevance calculates the features related to the conclusion via adding auxiliary components virtually. The evaluation function computes the contribution of adding auxiliary components through a particular strategy to the system solution.

## 3. Proposed Solution

Our method extends the traditional solver with (i) Monte-Carlo tree search balancing exploration and exploitation using the graph-attention mechanisms for estimating the prior probability of inferences to lead to the problem solving, and (ii) learning-based mechanisms for evaluation model.

### 3.1. RL framework for Adding Auxiliary Components

Since we apply Reinforcement Learning (RL) in the adding auxiliary components, we propose the basic elements of RL framework, including environment, state, action and reward.

**Environment**: In this task, the environment refers to the graph information and the whole problem information that processed by BERT. The environment retains consistent in the whole training process.

**State**: The state of an agent is concatenated with three parts: the embedding part from NLP results, the graphic configuration from Section 3.2 is restored to the original state, and a new reasoning library after adding auxiliary components. Known conditions and conclusions are used as the input. The selected auxiliary components are applied to the knowledge-based reasoning. If the conclusion proves to function, update the state space and exit.

**Action**: For the adding auxiliary components, an action refers to an agent choosing a strategy to step forward. Based on guidance value, it chooses the strategy according to the probabilities obtained by the model. Actions are either benefit or not to the final problem solving.

**Reward**: Reward is a feedback to the agent according to whether the action is benefit, and whether a series of actions can lead to the final solution in a specified number of times. If the conclusion proves to function by adding auxiliary component using the selected strategy, the evaluation model is updated. We give the reward 1 if it has benefit for the solving, otherwise, the reward is 0.

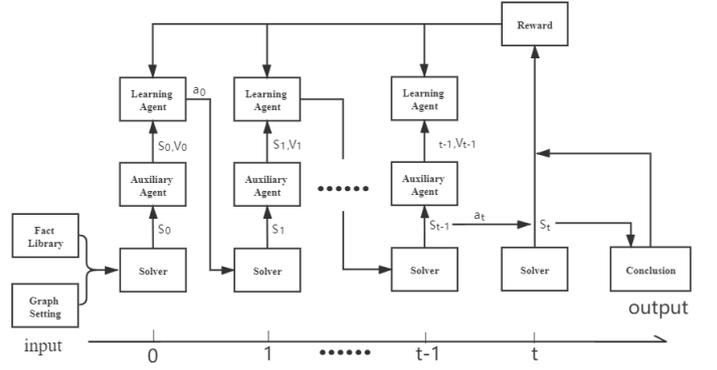

Figure 2: RL framework for A3C-RL

Figure 2 shows how a geometric problem is solved in our framework. Given a conjecture and a set of axioms, A3C-RL iteratively performs reasoning steps until a conclusion is proved (within a provided time limit). We define these four main elements by the tuple ($S, A, R, Q$), where $S$ is the set of states, $A$ is the set of actions, $R$ is the reward function, and $Q$ is the state transition probability. We further define $S, A, R$ and $P$ below. Let $s_t \in S$ denote the state at time $t$, and the set $S$ consists of all the possible values of $\{s_t, t \geq 0\}$ from the given graph information. Recalling that the system needs the entire history of traversed nodes and the query to make a correct decision, we define st by the following recursion:

$$s_t = s_{t-1} \cup \{a_t, v_t\} \quad (1)$$

where $a_t \in A$ denotes the action selected by the system at time $t$, $v_t \in V$ denotes the currently visited node that contains graph information after taking action $a_t$, and $V$ denotes the combination of the current graph $G$ with the auxiliary model library. Based on $s_{t-1}$, the system takes one of the following actions at each time $t$: (i) choosing a strategy in $A$ and moving to the next node, or (ii) terminating the searching, which is up to the max steps (=max_steps). The solver tracks the solving state, $s_t$, which encapsulates the clauses that have been derived or used in the derivation so far and the actions that can be taken by the solver at the current step. At each step, this state is passed to the learning agent - a deep learning model, discussed in Section 3.3, that predicts a distribution over the actions it uses to sample a corresponding action, $a_i$. This action is given to the solver, which executes it and updates the solving state.

In our experiment, we implement the MCTS with the standard UCT formula [13] to select the next actions with Equation 2. Once the STOP action is selected, the solver reaches the terminal state and outputs the result whether problem is solved. This is the transition probability of the action(inference) that leads from state $s_{t-1}$ to $s_t$. If no strategy learning is used, the prior probabilities ($q_t$) are all equal to one. The total reward for a node is computed as a sum of the rewards of all nodes below that node. In the basic setting, the reward for a leaf node is 1 if the sequence of inferences results in a closed tableau, i.e., the conclusion of the problem; Otherwise, it is 0.

$$R_{mcts} = \frac{w_t}{n_t} + c \cdot q_t \cdot \sqrt{\frac{lnN}{n_t}} \quad (2)$$

we maintain at each search node at time t the number of its visits $n_t$, the total reward $w_t$, its prior probability $q_t$, and N stands for the total number of visits of the parent node. The value of $c$ has been experimentally set to 2 when learned strategy selection. However, due to the searching time and performance issue, we improve selection function using BERT embedding to fit our problem, and we will introduce this model in Section 3.3.

*3.2. Graph Attention Strategy Network*

The knowledge-based library for generating strategies would be overwhelmed if we do not set up restrictions on the strategy network. To address this issue, we introduced graph attention mechanism using conclusion correlation $r(p, conc)$ between a point $p \in G$ and a conclusion conc in a problem to build a sub-graph. For different conclusions, it's better to focus more on points and lines which are highly related to the conclusions. The correlation is defined as,

$$r(p, conc) = \frac{1}{level(p)} + \frac{1}{2n}\sum_{i=1}^{2n}\frac{1}{level(p_i)} \quad (3)$$

Where $p_i$ is directly related to point $p$ in the previous level, and $n$ represents the number of points related to $p$. We designed an algorithm, called the Sub-graph Calculation Algorithm (SCA), shown in Algorithm 1, to build a sub-graph $G'$ from the original graph $G$ based on the correlation $r(p, conc)$. In SCA, *level(p)* is the level of point $p$, and *set(p')* is the set of all points in sub-graph $G'$. The codes from line 3 to line 15 select related points from $G$ to add to *set(p')* when the points meet the conditions. If $p$ is a point in the conclusion, we set *level(p)* = 1; otherwise, *level(p)* would be calculated based on its related point set *set(p'')* with a known condition that has a connection between $p$ and $p''$. If $p''$ is in the conclusion, then *level(p)* = min{ *level(p'')* + 1}. To determine how to select point $p$, we introduced a value *sign(p)*, defined as,

$$sign(p) = \begin{cases} 1, & r(p, conc) \geq \alpha \\ 0, & r(p. conc) < \alpha \end{cases} \quad (4)$$

Where α is the threshold set at 0.75 in with our experience. If $r(p, conc)$ is greater than or equal to α, *sign(p)* will be set to 1 and point $p$ will be added into *set(p')*; otherwise, *sign(p)* will be set to 0 and point p will be ignored.

As a fundamental step, the AttnStrategy needs to be built firstly based on the subgraph $G'$. We use the Drools rule inference engine[4] to generate a strategy network for adding auxiliary components. The knowledge-based factual library serves as the initial

---

[4]https://www.drools.org/

---

**Algorithm 1** The Sub-graph Calculation Algorithm

**Input Data**: an original graph $G$ and the conclusion, (*conc*), of a problem
**Output Result**: sub-graph $G'$
1: Let *level(p)* be the level of point $p$.
2: Let *set(p')* be the set of all points in sub-graph $G'$
3: **for** each point $p$ in $G$ **do**
4:   **if** $p$ is a point in *conc* **then**
5:     *level(p)* = 1;
6:   **else**
7:     Select a set of points related to $p$ according to conditions, noted as *set(p'')*;
8:     *level(p)* = *min*{*level(p'')*} + 1
9:   **end if**
10:   Calculate the correlation $r(p, conc)$ between $p$ and *conc* by formula 3;
11:   Calculate the value of sign(p) by formula 4;
12:   **if** 1 == *sign(p)* **then**
13:     *set(p')*.add(*p*);
14:   **end if**
15: **end for**
16: Calculate sub-graph $G'$ by *set(p')* and $G$.

---

factual condition for the engine. The generated auxiliary component knowledge under the state space can be defined as a model for adding auxiliary components, also called a rule.

In general, the AttnStrategy adds reasonable auxiliary components in accordance with current graphic information. For example, the AttnStrategy needs to verify connections between points, calculate coordinates of the auxiliary points, or check whether components are co-circular. In essence, the rule for adding auxiliary components consists of a set of common relations and graphical conditions. The graphical conditions that must be satisfied while adding auxiliary points include: 1) no collinear points; 2) no point coordinates and non-repeated point names.

The strategies can be represented as a set $A = a_t^m, 0 \leq m \leq L$, where $L$ is length of total strategies generated by Drool engine under the current state $s_t$. As shown in Table 1, before applying the AttnStrategy, there can be seven possible strategies in based on the state $s_t$. It would reduce to five strategies by AttnStrategy since the point $D$ would be ignore from the subgraph $G'$. Thus, the strategies with Code 2 and 4 were removed from the list, which is benefit for the final solution, illustrated on the following section.

*3.3. Training Evaluation Model by BERT*

In this section, we will introduce a deep learning model to select the top $m$ strategies based on AttnStrategy that can be used to solve the problem. There are altogether six main features, namely, segments equality in Formula 5, angles equality in Formula 6, lines parallel in Formula 7, lines perpendicular in Formula 8, congruent triangles in Formula 9, and similar triangles in Formula 10, to represent the graphs before and after adding auxiliary components.

| Code | Rules Generated by Strategy Network |
|------|-------------------------------------|
| 1 | Connect point $C$ and point $E$ |
| 2 | Connect point $B$ and point $D$ |
| 3 | Connect point $P$ and point $E$ |
| 4 | Connect point $E$ and point $D$ |
| 5 | Extend the midline of the triangle to make a parallelogram |
| 6 | Make the midpoint $F$ for $PB$ |
| 7 | Make the projection on the surface |

Table 1: Generated Strategy Network based on the Graph G for Example 1

$$f_{ij}^1 = \begin{cases} 1, & \frac{||i|-|j||}{\max\{|i|,|j|\}} \leq \beta \\ 0, & else \end{cases} \quad (5)$$

$$f_{ij}^2 = \begin{cases} 1, & \frac{|degree(i)-degree(j)|}{\max\{degree(i),degree(j)\}} \leq \beta \\ 0, & else \end{cases} \quad (6)$$

$$f_{ij}^3 = \begin{cases} 1, & 1 - \frac{degree(i,j)}{\pi} \leq \beta \\ 0, & else \end{cases} \quad (7)$$

$$f_{ij}^4 = \begin{cases} 1, & \frac{\left|\frac{\pi}{2}-degree(i,j)\right|}{\max\left\{\frac{\pi}{2},degree(i,j)\right\}} \leq \beta \\ 0, & else \end{cases} \quad (8)$$

$$f_{ij}^5 = \begin{cases} 1, & i \cong j \\ 0, & else \end{cases} \quad (9)$$

$$f_{ij}^6 = \begin{cases} 1, & i \sim j \\ 0, & else \end{cases} \quad (10)$$

$$v_t^k = \sum_{i=1}^n \sum_{j=2, i \neq j}^n f_{ij}^k, 1 \leq k \leq 6 \quad (11)$$

where parameters $i$ and $j$ represent two segments or two triangles in graph $G'$. $v_t^k$ in Equation 11 denotes the $k^{th}(1 \leq k \leq 6)$ of $v_t$, which accumulates the total features at time $t$ in the state $s_t$. The graph is often inaccurate, so we need to set up a tolerance limit $\beta$, also referred to as graph deviation hyperparameter, which in this paper was set at 0.13 based on our experimental experience. A function of the value network was employed to prune the strategy network. A strategy generated from the strategy network represents changes in features vectors before and after adding an auxiliary line component, which are calculated from the above characteristic formulas. The sets $V_{t-1} = (v_{t-1}^1, \cdots, v_{t-1}^6)$ and $V_t = (v_t^1, \cdots, v_t^6)$ represent vectors before and after adding auxiliary components by a strategy respectively. In example 1, $V_{t-1} = (5, 33, 9, 2, 1, 1)$ and different $V_t$ based on different strategies from Table 1, shown in the second column of Table 2. The combination of $V_{t-1}$ and $V_t$ are used as the input to our value network. If the strategy at is effective, the label of the value network is 1; otherwise, the label is 0.

We utilize the Bi-LTSM BiGRU model based on BERT embedding to train contribution value via collected dataset. For embedding, BERT settings were determined as described in[11], and we select sequence length = 64. For training, we select parameters as following: a batch size of 16, learning rate of 3e-5, and linear learning rate decay over 5 epochs. We concatenate $V_{t-1}$, $V_t$ as an input data, and predict the contribution value $F$ with 0 or 1 via Bi-LTSM_BERT model, shown in Equation 12.

| Code | $V_t$ | Evaluation $R$ | |
|------|-------|----------------|---|
| 1 | (5,42,10,2,1,1) | 235518 | (235518/236152)+0=0.9973 |
| 2 | (5,41,10,2,2,2) | 235518 | (235518/236152)+0=0.9973 |
| 3 | (6,39,9,2,1,1) | 235518 | (235518/236152)+0=0.9973 |
| 4 | (6,39,9,2,1,1) | 235518 | (235518/236152)+1=1.9973 |
| 5 | (9,62,13,4,4,3) | 99 | (99/236152)+1=1.0004 |
| 6 | (12,33,9,2,1,1) | 322 | (322/236152)+1=1.0014 |
| 7 | (17,82,12,3,3,3) | 213 | (213/236152)+1=1.0009 |

Table 2: Guidance Values based on Table 1

$$F = BiLTSM_BERT(V_{t-1}, V_t) \quad (12)$$

The output $R_{BERT}$ combines the contribution value $F$ of provided strategies with wining rate by calculating $w_t/n_t$, shown in Equation 13. The guidance value of each generated strategy for the example 1 shows in the last column of Table 2.

$$R_{BERT}(S_t) = \frac{w_t}{n_t} + F \quad (13)$$

We accumulate the $w_t$ into the database as evaluation function. As shown in Table 2, the evaluation function is listed in the third column, and it would apply into the Equation 13. For example, in Table 2, the top strategy – 'connect any point' has the highest value $w_t = 235518$, and the value of $n_t = 236152$, which shows it has totally simulated 236152 times for possible strategies. In this case, the strategy with Code 4 (R=1.9973) is selected, and it would be failed to solve the problem. But the strategy with Code 2 and 4 would be removed from strategy after applying AttnStrategy, the strategy with Code 6 is selected, and it can solve the problem successfully.

### 3.4. Automatically Adding Auxiliary Components based on Reinforcement Learning

This paper proposes a new algorithm, Automatically Adding Auxiliary Components based on Reinforcement Learning (A3C-RL), to assist in adding auxiliary components such as lines or points includes two main stages. In the first stage, a strategy network is generated based on AttnStrategy. The initial input for a strategy network includes current graphic information, knowledge-based fact library, and auxiliary component rule library which has been creat-ed. Drools inference engine generates reasonable strategies for adding auxiliary component based on initial input. In the second stage, adding auxiliary components based on RL framework, illustrated in Algorithm 2, is implemented.

---

**Algorithm 2** Training Algorithm of the Agent based for AttnStrategy and BERT

---

1: **for** episode=1 to N **do**
2:    Initialize state vector s0 using Equation 1 with AttnStrategy.
3:    Initialize num_steps to 0.
4:    **while** num_steps < max_steps **do**
5:      select a sample action $a_t \sim \pi_\theta(a_t|s_t)$.
6:      **if** Action solve the problem **then**
7:        $w_t = w_t + 1, n_t = n_t + 1$.
8:      **else**
9:        $w_t = w_t, n_t = n_t + 1$.
10:      **end if**
11:      Add $w_t$ to $W_{episode}$, $n_t$ to $N_{episode}$.
12:      Increment num_steps.
13:      **if** success or num_steps = max_steps **then**
14:        Update state space S .
15:        break.
16:      **end if**
17:    **end while**
18:    update $R_{BERT}$ with Equation 13 using $W_{episode}$ and $N_{episode}$.
19: **end for**

---

After the initializations, Line 5 in Algorithm 2 samples an action according to the output of the valuation network, where $\pi_\theta(a_t|s_t)$ denotes the probability of all strategies that calculates by Equation 13. The agent selects an action and obtains a reward. There are two main steps in the reasoning process.

1) Connect auxiliary lines: generating point collinear information, modifying line configuration, and trying to merge with known collinear information; otherwise, inserting new collinear information.

2) Make auxiliary points: generating a point, connecting auxiliary lines, and generating corresponding relationships according to types of auxiliary points, such as the mid-point, or the vertical point.

In Table 2, we can see that as Code 6 has the highest guidance value, A3C-RL would select the strategy with Code 6 to making a dynamic middle point F for segment PB. After A3C-RL is executed each time, the controller would restore the graphic configuration to the initial value, and rebuild a new reasoning library to start reasoning from the beginning. After it successfully reaches the conclusion or doesn't reach in a specified number of times, the rewards of the whole episode are used to update all parameters. The strategy learning data can be extracted from all search nodes or only from some of them. For value learning we characterize the proof states of the nodes by extracting features from all goals, the active path, and the whole tableau. Line 6 10 in Algorithm 2 assigned value $w_t = w_t + 1, n_t = n_t + 1$ if the problem is solved by taking action $a_t$; otherwise $w_t = w_t, n_t = n_t + 1$. Line 14 in Algorithm 2 updates the state space S with the search result, and Line 18 in Algorithm 2 updates guidance value $R_{BERT}$ using simulating results by accumulated value $W_{episode}$ and $N_{episode}$, where $W_{episode}$ and $N_{episode}$ represents the win steps and total steps after an episode.

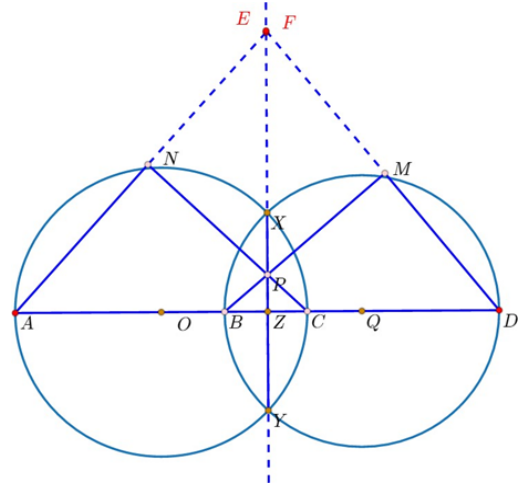

Figure 3: Complex Geometric Problem

## 4. Experiments

### 4.1. Experimental Setup

To test the proposed algorithm, we conducted an experiment. Database tools, such as MySQL and Hbase, were used for describing problems, storing pictures, and training samples. We trained the value network on a cluster that includes three servers each with 64G memory, 32 cores E5-2630 CPU, and 2TB storage. One server provided database service, NLP-TK, Kafka, and other services. The other two processed data input/output and human-like solving progress. On each of the three servers, a different GPU, namely, a dual 1080Ti, a dual 1080, and a single 1080, was installed and used for training the value network.

### 4.2. Dataset

We trained a sample set of 1000 geometric problems that cannot be solved without adding auxiliary components. We provided this sample set and the training dataset in the github[5] as supplementary materials. Each problem was approached by random selection of up to 10 auxiliary component strategies. To evaluate the effectiveness of generated strategies on adding auxiliary components, samples would be labelled as 1 or 0 for each strategy. The label was 1 for positive samples, and 0 for negative samples. A total of 43,152 labels were collected based on all generated strategies for 1000 geometric problems, including 8,427 positive and 34,725 negative labels. We attached this sample set as supplementary material. Combination vectors, representing the relationship between the graphs before and after addi-ing auxiliary components, are added as part of the input to the value network. We used the function F in formula 12 to measure the contribution of adding auxiliary components.

---

[5] https://github.com/GongqiLinVU/A3C-RL

| Algorithm | Accuracy (%) |
|---|---|
| A3C-RL (BiGRU) | 82.94 |
| A3C-RL (BiLTSM) | 83.50 |
| A3C-RL (CNN) | 80.36 |
| T-MCTS | 50.8 |

Table 3: Comparison of accuracy between A3C-RL and T-MCTS

*4.3. Success Rate in Adding Auxiliary Components*

We utilize the BERT-base model on our collected dataset for the vector embedding. For comparison, three different classification models were used for training: BiLTSM [24], Bidirectional Gated Recurrent Unit (BiGRU), and CNN[25]. We took the same proportion of training/valid/test set by 80%/10%/10% on three models, and the proportion were was randomly selected from 8,427 positive and 34,725 negative samples. As shown in Table 3, The accuracy of A3C-RL with BiLTSM has the best result, up to 83.50%, which is 32.7% more accuracy than the traditional MCTS methods (T-MCTS) with Equation 2 that only hit 50.8% accuracy.

*4.4. Effects of RL Framework*

We designed and implemented two experiments to assess the performance of A3C-RL. We in this section used Figure 4 (Example 2) as a single case to illustrate its complexity, and applied A3C-RL on a batch data set with 9,939 geometric problems to show its efficiency. Furthermore, a comparison between A3C-RL and humans was made under real test environment where geometric problems were randomly selected from the University Entrance Examination of China.

*4.4.1. Single Case Experiment*

To illustrate how to solve complex geometric problems, we used a single case of a geometric question from the Olympic competition, which is usually much more difficult than most of the questions in the annual University Entrance Examination of China, shown in Figure 3. As knowledge from expert experiences, the problem in Figure 3 cannot be solved without adding auxiliary components. For a fair comparison, this problem should be solved using the strategy network within a reasonable allowable length of time, such as 30 minutes, which we selected for this experiment.

The geometric question in Figure 3 is that "Suppose point *A,B,C* and *D* are four different points arranged in turn on a straight line, the line intersects with the circle *O* having a diameter *AC* at point *X*, and intersects with the circle *Q* having a diameter *BD* at point *Y*. Line *XY* intersects *BC* with point *Z*, if point *P* is a point different from *Z* on line *XY*, the line *CP* intersects with the circle *O* having a diameter *AC* at point *C* and *M*, the line *BP* intersects with the circle *Q* having a diameter *BD* at point *B* and *N*. Prove: Lines *AM*, *XY* and *DN* intersect at one point.".

This complex question manifested the contribution of ranking strategies by value network. The problem cannot be solved by applying the strategy network within a reasonable length of time alone. As shown in Table 4, there are a total of 3,756 strategies generated from the strategy net-work, including some extremely time-consuming strategies, such as Strategy 1213. We selected top 10 value network strategies as auto-solving strategies, shown in Table 5. Based on our trained models, No. 4 and No. 5 strategies proved to be effective with their highest values of *R*, calculated through cross-validation. We attached the whole problem-solving progress as supplementary material in our github[6].

| 1 | Connect point *M* and point *O* |
|---|---|
| 2 | Connect point *N* and point *O* |
| 3 | Create middle point *G* of segment *AM* |
| ...... | ...... |
| 1213 | Extended segment *DN* intersection segment *XY* at point $X_{107}$ |
| ...... | ...... |
| 1625 | Extended segment *AM* intersection segment *XY* at point $X_{155}$ |
| ...... | ...... |
| 3756 | Connect point $X_{314}$ and point $X_{352}$ |

Table 4: Generating 3756 adding auxiliary line strategies based on Strategy Network

| 1 | Create middle point *G* of segment *DN*, connect point *G* and point *Q* |
|---|---|
| 2 | Create middle point *G* of segment *AM*, connect point *G* and point *O* |
| 3 | Connect point *X* and point *O* |
| 4 | Extended segment *AM* intersection segment *XY* at point *E* |
| 5 | Extended segment *DN* intersection segment *XY* at point *F* |
| 6 | Create vertical segment MG of segment XY through point M which the foot is point G |
| 7 | Create vertical segment AG of segment DN through point A which the foot is point G |
| 8 | Extended segment AM intersection segment DN at point G |
| 9 | Connect point *M* and point *N* |
| 10 | Connect point *N* and point *Q* |

Table 5: The selected top 10 candidates as the branching auto solving strategies.

*4.4.2. Batches Experiment*

We verified A3C-RL on a batch set with 9,939 geometry problems written in Chinese. There were three main steps for this batches experiment. In the first step, 9,939 geometry problems were solved using traditional logical reasoning and computational reasoning directly. Then, the corresponding training sets were generated from the results from the first step, including 800,000 relations, 46,897 state spaces, 161,715 training samples, and 26,321 graphic feature vectors before and after adding auxiliary components using formulas 5-11. Finally, the value network was generated from the training set provided in the Data Set session by following the A3C-RL workflow in Figure 3, and the 9,939 problems were solved again. Before using the A3C-RL algorithm, the accuracy rate was only 68.7%. It reached 80.3% after the new algorithm was applied.

---
[6]https://github.com/GongqiLinVU/A3C-RL

### 4.5. Practical Experiment

To verify the performance of the A3C-RL algorithm, we randomly selected ten questions from the University Entrance Mathematical Examination of past years, provided by iFly zhixue[7]. All these ten questions contained 2 or 3 sub questions. We provided this question set and the generated human-like solutions in the github[8] as supplementary materials. Senior high-school students, randomly selected from different areas in China, answered the ten questions. We used average accuracy rate (AAR) to measure how accurately the selected students answered the questions, which can be calculated as following:

$$AAR = \frac{\sum_{i=1}^{P_{num}} a(i)}{P_{num}} \quad (14)$$

where $a(i)$ represents the rate of $i^{th}$ participant, and $p_{num}$ is the total number of participants on a selected question. AAR not only represents the human performance, but also demonstrates the difficulty level of a selected question. For A3C-RL, we only recorded whether it was solved or not, and how many sub questions were solved, e.g., All Solved (3/3) means all three sub-questions are solved, and Partially Solved (2/3) represents that only two sub questions are solved in total three sub questions. The results are reported in Table 6. The first column in Table 6 represents the number of attended senior high school students for each question. The AAR of humans and the result of A3C-RL are shown in the last two columns. From Table 6, we can find that A3C-RL not only performs well on the problems that human can solve well, but also it can be done very well on some questions that human does not that well. For example, the AARs of question 1, 8, 9 and 10 are under 50%, but A3C-RL can perfectly solve the three questions, especially for question 9, which only 29.06% students can solve this question. However, A3C-RL still failed on some sub questions, such as question 4, 7 or 8, since it selected the incorrect strategies.

### 5. Conclusion and Future Work

In this paper, we propose AttnStrategy to reduce the strategy searching space by graph attention mechanism for solving geometric problems. Based on the language model, We construct the deep reinforcement learning framework by incorporating wi-th AttnStrategy from traditional machine proofs, called A3C-RL algorithm. The human-like problem solving processes are automatically generated based on A3C-RL algorithm, and we improve the solving process by predicting the strategy with training evaluation. At present, the training set contains approximately 9,939 geometric problems. With additional data input, we can continue to improve the ability of solving problems by training and learning. There are still some research directions warranting further exploration. One of the directions is how to deal with a large number of strategy branches simultaneously. We applied A3C-RL algorithm to generate a humanlike response system in this paper. However, for a large number of strategy branches in logical reasoning and floating point operations, we will try to consider introducing parallel operations to improve the efficiency of the system in the future. We also consider applying A3C-RL framework on automated mathematical reasoning that provides interactive and cognitive learning modes for online education. This online interactive learning mode provides students with clear problem-solving mind maps and improves their ability to solve problems.

---
[7]https://www.zhixue.com/
[8]https://github.com/GongqiLinVU/A3C-RL

| No. | Participants | AAR by humans (%) | Result by A3C-RL |
|-----|--------------|-------------------|-------------------|
| 1 | 662 | 49.58 | All Solved (3/3) |
| 2 | 175 | 64.62 | All Solved (2/2) |
| 3 | 229 | 84.40 | All Solved (3/3) |
| 4 | 272 | 45.16 | Partially Solved (1/2) |
| 5 | 267 | 88.56 | All Solved (3/3) |
| 6 | 30657 | 80.93 | All Solved (3/3) |
| 7 | 2502 | 59.76 | Partially Solved (2/3) |
| 8 | 444 | 43.92 | Partially Solved (1/3) |
| 9 | 464 | 29.06 | All Solved (3/3) |
| 10 | 911 | 45.44 | All Solved (3/3) |

Table 6: AAC-KVN vs. Human on AAR

**Acknowledgements**

The authors wish to thank Philip Hamish Todd of Saltire Software, Inc, Shengchuan Wu of Franz corporation, Xinchao Wu and other students in our lab. The authors also wish to thank the anonymous reviewers for their helpful comments. This work was funded by the National Key R&D Program of China (No. 2018YFB1005100 & No. 2018YFB1005104), the National Natural Science Foundation of China (No. 61876034, 61202257, 61650110512), the China Postdoctoral Science Foundation (No. 2016M602677) and the Science and Technology Incubation and Achievement Transformation Project of Neijiang City, Sichuan Province, China (No. 2019KJFH005).

**References**

[1] A. J. Robinson, A. Voronkov, Handbook of automated reasoning, Vol. 1, Gulf Professional Publishing, 2001.

[2] G. Cheng, W. Zhu, Z. Wang, J. Chen, Y. Qu, Taking up the gaokao challenge: An information retrieval approach., in: IJCAI, 2016, pp. 2479–2485.

[3] P. Clark, Elementary school science and math tests as a driver for ai: take the aristo challenge!, in: Twenty-Seventh IAAI Conference, 2015.

[4] A. Naumowicz, A. Korniłowicz, A brief overview of mizar, in: International Conference on Theorem Proving in Higher Order Logics, Springer, 2009, pp. 67–72.

[5] J. Harrison, Hol light: An overview, in: International Conference on Theorem Proving in Higher Order Logics, Springer, 2009, pp. 60-66.

[6] G. Irving, C. Szegedy, A. A. Alemi, N. Eén, F. Chollet, J. Urban, Deepmath-deep sequence models for premise selection, Advances in neural information processing systems 29 (2016) 2235–2243.

[7] D. Huang, P. Dhariwal, D. Song, I. Sutskever, Gamepad: A learning environment for theorem proving, arXiv preprint arXiv:1806.00608 (2018).

[8] K. Bansal, S. Loos, M. Rabe, C. Szegedy, S. Wilcox, Holist: An environment for machine learning of higher order logic theorem proving, in: International Conference on Machine Learning, 2019, pp. 454–463.

[9] A. Paliwal, S. M. Loos, M. N. Rabe, K. Bansal, C. Szegedy, Graph representations for higher-order logic and theorem proving., in: AAAI, 2020, pp. 2967–2974.

[10] D. Lee, C. Szegedy, M. N. Rabe, S. M. Loos, K. Bansal, Mathematical reasoning in latent space, arXiv preprint arXiv:1909.11851 (2019).

[11] J. Devlin, M.-W. Chang, K. Lee, K. Toutanova, Bert: Pre-training of deep bidirectional transformers for language understanding, arXiv preprint arXiv:1810.04805 (2018).

[12] A. Radford, J. Wu, Rewon child, david luan, dario amodei, and ilya sutskever. 2019, Language models are unsupervised multitask learners (2019).

[13] L. Kocsis, C. Szepesvári, Bandit based monte-carlo planning, in: European conference on machine learning, Springer, 2006, pp. 282–293.

[14] R. Sutton, anda. g. barto, reinforcement learningan introduction (1998).

[15] A. Vaswani, N. Shazeer, N. Parmar, J. Uszkoreit, L. Jones, A. N. Gomez, Ł. Kaiser, I. Polosukhin, Attention is all you need, Advances in neural information processing systems 30 (2017) 5998–6008.

[16] A. Radford, K. Narasimhan, T. Salimans, I. Sutskever, Improving language understanding by generative pre-training (2018).

[17] T. B. Brown, B. Mann, N. Ryder, M. Subbiah, J. Kaplan, P. Dhariwal, A. Neelakantan, P. Shyam, G. Sastry, A. Askell, et al., Language models are few-shot learners, arXiv preprint arXiv:2005.14165 (2020).

[18] C. Raffel, N. Shazeer, A. Roberts, K. Lee, S. Narang, M. Matena, Y. Zhou, W. Li, P. J. Liu, Exploring the limits of transfer learning with a unified text-to-text transformer, arXiv preprint arXiv:1910.10683 (2019).

[19] W. Ling, D. Yogatama, C. Dyer, P. Blunsom, Program induction by rationale generation: Learning to solve and explain algebraic word problems, arXiv preprint arXiv:1705.04146 (2017).

[20] A. Amini, S. Gabriel, P. Lin, R. Koncel-Kedziorski, Y. Choi, H. Ha-jishirzi, Mathqa: Towards interpretable math word problem solving with operation-based formalisms, arXiv preprint arXiv:1905.13319 (2019).

[21] G. Lample, F. Charton, Deep learning for symbolic mathematics, arXiv preprint arXiv:1912.01412 (2019).

[22] M. Dehghani, S. Gouws, O. Vinyals, J. Uszkoreit, Ł. Kaiser, Universal transformers, arXiv preprint arXiv:1807.03819 (2018).

[23] D. Saxton, E. Grefenstette, F. Hill, P. Kohli, Analysing mathematical reasoning abilities of neural models, arXiv preprint arXiv:1904.01557 (2019).

[24] G. Liu, J. Guo, Bidirectional lstm with attention mechanism and convolutional layer for text classification, Neurocomputing 337 (2019) 325–338.

[25] Y. Kim, Convolutional neural networks for sentence classification, arXiv preprint arXiv:1408.5882 (2014).